\documentclass[conference]{IEEEtran}
\pdfoutput=1
\IEEEoverridecommandlockouts
\usepackage{cite}
\usepackage{amsmath,amssymb,amsfonts}
\usepackage{algorithmic}
\usepackage{graphicx}
\usepackage{textcomp}
\usepackage{xcolor}

\usepackage{url}
\usepackage{makecell,booktabs}

\usepackage{xcolor}

\DeclareMathOperator{\argmin}{arg\,min}

\def\BibTeX{{\rm B\kern-.05em{\sc i\kern-.025em b}\kern-.08em
    T\kern-.1667em\lower.7ex\hbox{E}\kern-.125emX}}
\begin{document}

\title{Handling Structural Mismatches in Real-time \\ Opera Tracking\\
\thanks{The research is supported by the European Union under the EU's Horizon 2020 research and innovation programme, Marie Sk\l{}odowska-Curie grant agreement No.~765068.
The LIT AI Lab is supported by the Federal State of Upper Austria.}
}

\author{\IEEEauthorblockN{Charles Brazier}
\IEEEauthorblockA{\textit{Institute of Computational Perception} \\
\textit{Johannes Kepler University}\\
Linz, Austria\\
charles.brazier@jku.at}
\and
\IEEEauthorblockN{Gerhard Widmer}
\IEEEauthorblockA{\textit{Institute of Computational Perception and} \\
\textit{LIT AI Lab, Linz Institute of Technology} \\
\textit{Johannes Kepler University}, Linz, Austria\\
gerhard.widmer@jku.at}
}

\maketitle

\begin{abstract}
Algorithms for reliable real-time score following in live opera promise a lot of useful applications such as automatic subtitles display, or real-time video cutting in live streaming. Until now, such systems were based on the strong assumption that an opera performance follows the structure of the score linearly. However, this is rarely the case in practice, because of different opera versions and directors' cutting choices. In this paper, we propose a two-level solution to this problem. We introduce a real-time-capable, high-resolution (HR) tracker that can handle jumps or repetitions at specific locations provided to it. We then combine this with an additional low-resolution (LR) tracker that can handle all sorts of mismatches that can occur at any time, with some imprecision, and can re-direct the HR tracker if the latter is `lost' in the score. We show that the combination of the two improves tracking robustness in the presence of strong structural mismatches.
\end{abstract}


\section{Introduction}
\label{sec:introduction}

Real-time score following aims at synchronizing a live music performance (audio stream) with the corresponding sheet music in order to identify and track the current score position over time. Previous works in the domain have led to a variety of useful applications such as automatic page-turning for pianists \cite{arzt2008automatic}, live performance visualization \cite{lartillot2020real}, or score viewing and automatic contextualization in orchestra concerts \cite{prockup2013orchestral} \cite{arzt2015artificial}. Such a system would also be useful in opera houses, with applications such as automatically synchronized subtitles in the hall, or automated camera selection and real-time video cutting in live streaming.

Opera tracking is particularly challenging, due to the complexity and partial unpredictability of acoustic events in this genre, and it is only recently that there has been some initial work in this direction \cite{brazier2020towards} \cite{brazier2020addressing}. In addition, machine-readable symbolic scores are generally not available for entire operas, due to the tedious manual work involved. To address this, the basic approach is to perform audio-to-audio alignment between the live performance (the \textit{target}) and another recording of the same opera (the \textit{reference} performance; possibly a commercial recording by different artists), that has been marked up with timestamps at relevant places. The idea is that the reference recording serves as a proxy to the score. In \cite{brazier2020towards}, the authors combined an On-Line Time Warping (OLTW) audio-to-audio matching algorithm with three acoustic classifiers (for applause, music, and speech detection) that can directly control the tracker in cases of spontaneous applause, stage noises, interludes, or breaks during the performance. \cite{brazier2020addressing} extended this to using two OLTW-based trackers working in parallel, one focusing on music-, the other on speech-sensitive features, to improve tracking, especially during \textit{recitativo} sections.

Previous attempts at opera tracking make the strong assumption that reference and target performances follow the same musical structure. However, this assumption is often violated, for a number of reasons. First, an opera may exist in multiple versions where each version replaces or inserts specific parts (which are sometimes listed in an appendix to the score book). Second, a live opera can include parts that are not written in the score at all, such as cadenzas or interludes that are spontaneously played between parts, mostly to accompany the actors getting ready on stage. Finally, a director may decide to stage a shortened version of the work, omitting certain songs or parts, or repetitions inside parts. Sometimes [as reported by our project partner, the Vienna State Opera] it even happens that an aria is spontaneously repeated in response to enthusiastic cheering from the audience.

To cope with such structural mismatches between reference and target in real time, we first describe a high-resolution (HR) tracker, based on a new adaptive version of the JumpDTW algorithm \cite{fremerey2010handling}, that can handle small jumps or repetitions at specific locations. We then propose an additional low-resolution (LR) tracker, inspired by \cite{arzt2010towards} and \cite{shan2020improved}, that can handle all sorts of mismatches (forward or backward jumps) that can occur at any time, by considering all score positions as possible candidates (at a low temporal resolution). Finally, we describe how to integrate these two by permitting the LR tracker to re-direct the HR tracker when the two do not match and the LR tracker is highly confident. We will show that this two-level strategy improves tracking robustness in the presence of strong structural mismatches.


\section{Related Work}
\label{sec:litterature}

There are several existing works that tackle the structural mismatch problem in the context of classical music. In the \textit{offline case}, M\"uller and Appelt \cite{muller2008path} construct the alignment path by first enhancing partial paths that better reveal global structural aspects in the similarity matrix. Grachten et al. \cite{grachten2013automatic} propose a variant of the Dynamic Time Warping (DTW) algorithm that can skip parts that seem dissimilar. The two techniques have the advantage to not require explicit information of jump locations; they cannot handle repetitions, however. \textit{JumpDTW} \cite{fremerey2010handling} extends the DTW algorithm by connecting the beginnings and ends of all parts where jumps can occur; this requires us to know the position of the score blocks beforehand. In \cite{shan2020improved}, Shan and Tsai describe a hierarchical DTW algorithm that first performs Subsequence DTW \cite{muller2015fundamentals} between each sheet music line and the entire recording, and then links segments via a JumpDTW variant. This method can handle jumps and repeats when jump locations are unknown; however, it cannot be used in an on-line, real-time context.

In the case of \textit{online} alignment, Arzt et al. \cite{arzt2008automatic} compute three alignment hypotheses in parallel, related to the repeat, continue or skip actions. In \cite{arzt2010towards}, Arzt and Widmer propose a two-level hypothesis tracking process, where a higher-level tracker evaluates alternative possible positions in the score and feeds the best candidates to a lower-level tracker that applies standard OLTW. Finally, in \cite{nakamura2015real}, Nakamura et al. use a HMM variant that connects all score events together with additional repeat and skip probabilities.

The method proposed in this paper works in real time. A first tracker (HR) handles repeats and skips that involve a small number of skipped parts, reflecting the most common mismatches occurring in real conditions. A second tracker (LR) handles all other eccentric structural mismatches (including forward and backward jumps, with an unlimited amount of skipped parts) that can occur at any time. The HR tracker's score position is reset to the LR position when the two trackers disagree and the LR reliability factor is maximal.

\section{Data description}
\label{sec:data}

This study focuses on the opera \textit{``Don Giovanni"} by W.A.Mozart and makes use of a dataset described in \cite{brazier2020addressing}, which comprises three different performances.\footnote{We would have liked to compile a larger annotated dataset, but even precisely annotating three full opera recordings (500 score pages, 8 hours of audio, 16,000 bar annotations overall) already took around 300 person hours. Additional operas are currently being worked on.} The first one, to serve as \textit{reference} and proxy to the printed score, is a CD recording (Deutsche Grammophon) of a live performance conducted by Herbert von Karajan in 1985. The recording was manually annotated at the bar level. In the end, the score book of more than 500~pages has been linked to the recording with 5,320~bar annotations (2,877 for the first act, 2,443 for the second act). As \textit{target} performances to be used for live tracking tests, we use two live performances recorded at the Vienna State Opera\footnote{\url{https://www.wiener-staatsoper.at/en/}}, our partner in this project, conducted by \'Adam~Fischer (2018) and Antonello~Manacorda (2019), respectively (two different productions with different casts). Each performance involves 5,304 manual bar annotations that will serve to evaluate the tracker.

The score of the opera was provided by the \textit{Salzburg Mozarteum Foundation}, and its structure is detailed in a table available online\footnote{\url{dme.mozarteum.at/DME/nma/nma_toc.php?vsep=68}}. In this table, we consider each new title as a new part of the opera. Thus, the first act is composed of 55 parts (from Ouvertura - Andante to No.~13: Finale (\textit{Sul tuo capo in questo giorno})). The second act is composed of 39 parts (from No.~14: Duetto (\textit{Eh via buffone}) to No.~24 Finale - Presto (\textit{Questo {\`e} il fin di chi fa mal})). The score also includes an annex with 9 parts pertaining to an alternative version (the \textit{Vienna version)} of the opera. With two exceptions, the beginnings and ends of the parts correspond to the beginnings and ends of bars. Thus the indexes of part transitions are known without additional annotations.

The performances we have in our dataset have a very similar structure. The two target performances are identical. They play the same annex parts as the reference, but both exclude one part in the first act (\textit{In questa forma dunque}, 11 bar annotations), and one in the second (\textit{Ah, si segua il suo passo}, 5 bar annotations). Since those structural mismatches are not sufficient for a systematic experimental evaluation, we simulated random skips and repetitions in the two performances. Thus, our evaluation dataset is composed of 10 opera versions of each of the two target performances, where for each version, between 1/3 and 2/3 of the parts are randomly removed, and where one of the parts that are followed by an applause sequence is repeated. Spontaneous applause, interruptions, and interludes that sometimes appear in between the parts are kept in the different target versions. In this way, the experimental evaluation will test the tracking robustness to random `forward jumps', `repetitions', and `inserted parts', representing the most probable cases that can occur during a live performance. In practice, `backward jumps' other than repetitions cannot happen but they can frequently occur during a rehearsal. There is nothing in our model that would prevent it from handling this specific case.

\section{Model}
\label{sec:model}

As mentioned before, score following is realized by computing, in real time, an audio-to-audio alignment between the incoming target performance and a reference performance. From the audio files, we first extract MFCC features \cite{gadermaier2019study} with a window size of 20~ms and a hop size of 10~ms. The features in the \textit{reference} are computed beforehand, giving a sequence of 559,038 frames for the first act and 508,849 frames for the second. The features in the target are computed in real time. Then for each new incoming target feature vector, we compute the cosine distance between the feature and an interval $c$ of reference features centered around the expected score position, which gives us a linear time complexity (in practice, ${c=4000}$, corresponding to a context of 40~seconds of audio).

\subsection{Baseline Algorithm}
\label{sec:baseline}

As a baseline, we use the algorithm detailed in \cite{brazier2020towards}; it works as follows. Given the reference feature sequence ${X = x_{1}, ..., x_{M}}$, the new target feature $y_{j}$, the accumulated cost vector from the previous target feature $D_{j-1}$ and its previous score position
${sp_{j-1} = \argmin_{i\in \left[1:M\right]} D_{j-1}}$,
the new accumulated cost vector $D_{j-1}$ is computed as

$\forall i\in \left[ sp_{j-1}-c/2:sp_{j-1}+c/2\right],$
\begin{equation}
D_{j}[i] = \rm{cdist}(x_{i}, y_{j}) + \min \begin{cases}
        D_{j-1}[i-1]\\
        D_{j-1}[i]\\
        D_{j}[i-1]\\
    \end{cases}
\label{eqn:classicalDTW}
\end{equation}

The recursive formula shows that each cumulative cost cell, representing a unique score position, is linked to itself and its predecessor. This results in an alignment path that includes all intermediate score positions between the first and the last score positions given by the alignment, without the ability to skip or repeat parts. In the following, we first propose to extend the recursive formula to handle jumps or repetitions at known locations (Section~\ref{sec:HRTracker}). We then propose the use of an additional tracker, working in parallel, that can handle spontaneous jumps anytime with robustness but at the expense of accuracy (Section~\ref{sec:LRTracker}).

\subsection{High-Resolution Tracker: Jump On-Line DTW}
\label{sec:HRTracker}

The purpose of the HR tracker is to handle backward and forward score jumps or repetitions at known locations. This represents the most common real case. For computational reasons, it will only handle a small number of jump hypotheses (more extreme cases will be handled in Section~\ref{sec:LRTracker}). It adapts the JumpDTW \cite{fremerey2010handling} algorithm to its forward path only, and operates at the feature level.

To handle score jumps that appear at specific jump locations, Fremerey et al. \cite{fremerey2010handling} proposed to add connections in the recursive formula. Let $K$ be the total number of parts in the opera. Following the notation in \cite{fremerey2010handling}, let ${S:=\left\{ s_{k} ~|~ k\in \left[ 1:K\right] \right\}}$ and ${T:=\left\{ t_{k} ~|~ k\in \left[ 1:K\right] \right\}}$ be the list of all score indexes that represent, respectively, the start $s_{k}$ and end $t_{k}$ of a part ${k\in K}$. The Jump On-Line Time Warping applies Equation~\eqref{eqn:classicalDTW} whenever ${i\notin S}$. For all ${i\in S}$ (i.e., at possible jump points),
\begin{equation}
D_{j}[i] = \rm{cdist}(x_{i}, y_{j}) + \min \begin{cases}
        D_{j-1}[i-1]\\
        D_{j-1}[i]\\
        D_{j}[i-1]\\
        D_{j-1}[t], \forall t \in T\\
    \end{cases}
\label{eqn:jumpDTW}
\end{equation}

The equation shows that each score position that corresponds to a start of a part is linked to its previous score position but also to all positions ${t\in T}$ that correspond to an end of a part. For a part ${k\in \left[ 1:K\right]}$, the transition from $t_{k}$ to $s_{k_{b}}$ with $k>1$ and ${k_{b}}<k$ corresponds to a backward jump, the transition from $t_{k}$ to $s_{k}$ to a repetition, the transition from $t_{k}$ to $s_{k+1}$ with $k<K$ corresponds to continuing, and the transition from $t_{k}$ to $s_{k_{s}}$ with $k<K-1$ and ${{k_{s}}>k+1}$ corresponds to a skip in the score.

Adding new connections to the recursive formula increases the size of the calculation vector. To ensure a real-time computation, we limit the calculation to 8 possible part transitions: repetition of the current part; regular continuation; or jumps to the next 6 parts in the score. This multiple hypotheses tracking starts naturally when the score position is higher than the value ${t_{k}-c/2}$, $k$ being the current tracking part. As soon as the current score position given by the tracker is not in ${\left[ t_{k}-c/2:t_{k}\right]}$ and not in ${\left[ s_{k^{*}}:s_{k^{*}}+500\right], \forall k^{*}\in \left[ 1:K\right]}$, we consider that a new part is detected and we apply Equation~\eqref{eqn:classicalDTW} at the next frame.

\subsection{Low-Resolution Tracker}
\label{sec:LRTracker}

The purpose of our second component, the LR tracker, is to ensure robust tracking in case of arbitrary mismatches, including jumps from and to arbitrary places. This tracker considers all score positions as hypothetical candidates. It cannot operate at the feature level and thus works at a low resolution. It interacts with the HR tracker only if the two trackers' outputs do not match and the current LR position hypothesis is considered reliable. 

The LR tracker is not constrained to a limited score window but can consider all possible score positions during tracking. As it is impossible to handle the calculation of more than 500,000~distances per frame, it works at a considerably lower resolution. The low-resolution features are computed from the high-resolution features in following the method presented in \cite{arzt2010towards}, by convolving the original feature sequence with a Hann window of length 600~ms and a hop size of 300~ms. Thus, the score feature sequence is reduced to 18,652~frames for the first act and 16,979~frames for the second act. The part start and end indexes are listed in $S_{LR}$ and $T_{LR}$.

The LR tracker works as follows. For each new incoming LR feature, we compute the cosine distance between this and all the LR score features. Then, keeping in memory the 30 last distance vectors in a cost matrix $C$, we compute recursively the cumulative cost vector $D_{30}$ with the help of the diagonal matching function (see \cite{muller2015fundamentals}) where jumps are allowed. We initialize the cumulative matrix $D_{1}$ by $C_{1}$, allowing the alignment to start at any point. Then,

$\forall l\in \left[ 2:30\right],\forall i\in \left[ 1:N_{LR}\right],$
\begin{equation}
D_{l}[i] = C_{l}[i] + \begin{cases} +\infty \text{, if $i=0$}\\
                                    D_{l-1}[i-1] \text{, if ${i\notin S_{LR}}$}\\
                                    \min (D_{l-1}[i-1], D_{l-1}[t]~\forall t \in T_{LR})  \text{, else}\\
                                    \end{cases}
\label{eqn:jumpdiagonal}
\end{equation}

The use of the LR tracker differs from the strategy described in \cite{arzt2010towards}, where the authors showed that the estimator was robust enough to output an ordered list of possible current score positions, where the `best' top 5 hypotheses were reliable. However, this fails in the case of opera, due to the lack of harmonic similarity in some passages (e.g., during recitatives). To increase robustness, we aim at connecting the output vectors $D_{30}$ through time, retrieving a hypothetical position, and proposing a reliability factor based on the last 30 hypothetical positions. This proposal has a strong correlation with \cite{shan2020improved}. After the calculation of the previous $D_{30}$ corresponding to the subsequent alignment between the last 30 LR target frames and the whole reference audio, we update a cumulative cost matrix with the JOTLW algorithm which applies Equation \eqref{eqn:jumpDTW} in considering the output $D_{30}$ as the new cost. This aims at linking the segments between themselves. Then we extract the minimum value $x_{LR_j}$ of the cumulative cost matrix, $j$ being the current index in the target sequence, and consider it as a hypothetical score position. Considering ${\Delta_{x}:= j \mapsto x_{j} - x_{j-30}}$, if the last 30~values of $\Delta_{x_{LR}}$ are between 15 and 45, we set our reliability factor to $\mathit{rf}=1$. Else, $\mathit{rf}=0$. That is, if the last 30 score positions can be approximated by a line with a slope $a\in [0.5, 1.5]$, we are confident about the LR score position. Musically speaking, this means that the tracker has a strong belief in aligning two times series with tempo variations included in the previous interval.

\begin{figure}[t]
\centering
\includegraphics[width=0.9\columnwidth]{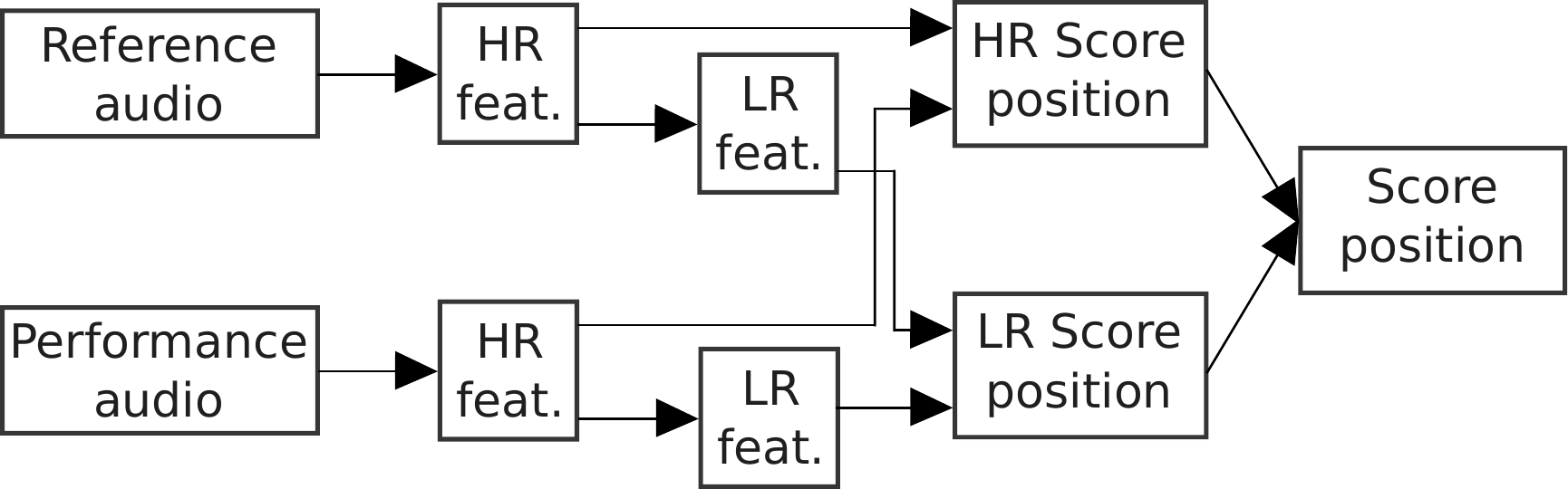}
\caption{Integrated model. Live Performance features are extracted in real-time, while reference features are extracted beforehand.\label{fig:model}}
\end{figure}

The LR tracker interacts with the HR tracker only if the reliability factor $\mathit{rf}$ is equal to 1. If $\mathit{rf}=0$, we consider the LR score position as the final score position. If $\mathit{rf}=1$ and the HR score position is not in the interval given by the LR tracker, we consider the middle of this interval as the final score position and reset the HR cumulative score matrix values to the value of the LR cumulative score matrix at the LR score position. The other values are set to $+\infty$. In this case, the LR tracker re-directs the HR tracker to its belief.
The final tracking process is resumed in Figure~\ref{fig:model}.

\section{Experiments and Discussion}
\label{sec:experiments}

In the following experiments, we evaluate the accuracy of different trackers on our augmented opera dataset (see Section~\ref{sec:data}). For each tracker, we compute the (on-line) alignments between the 20 randomly modified opera versions and the full reference. Each alignment gives a precise linkage between the target feature sequence and the reference feature sequence. For each target frame, we retrieve the corresponding part and bar ids. If the part/bar id matches to its ground truth extracted from the manual annotations, we consider it as a match. We also provide the alignment accuracy that corresponds to an acceptable error up to 5 bars.

\subsection{Evaluation of High-Resolution Trackers}
\label{sec:eval_HR}

In a first experiment, we compare four different models that can work at the frame level. The first one, \textit{Baseline}, refers to \cite{brazier2020towards} and uses the baseline algorithm (Section~\ref{sec:baseline}). The second model, \textit{JOLTW}, uses the HR tracker only (Section~\ref{sec:HRTracker}). \textit{Baseline+LR tracker} employs two-level tracking, combining \textit{Baseline} with the LR tracker. Finally, \textit{JOLTW+LR tracker} uses LR and HR trackers. In total, the dataset involves 12 million frames. The results are given in Table~\ref{tab:experiments}. 

\begin{table} [t]
\small
 \begin{center}
 \scalebox{1}{ 
 \begin{tabular}{@{}lccc@{}}
  \toprule
  \thead{\textbf{{\normalsize Model}}}
    &   \thead{\textbf{{\normalsize Part-wise}}\\ \textbf{{\normalsize align. acc.}}}
    &   \thead{\textbf{{\normalsize Bar-wise}}\\ \textbf{{\normalsize align. acc.}}}
    &   \thead{\textbf{\normalsize {@5 bars}}\\ \textbf{{\normalsize align. acc.}}}\\
  \midrule
  \textbf{Baseline} & 10.4\%  & 7.2\% & 9.2\%\\
  \textbf{JOLTW} & 28.3\%  & 17.3\% & 21.7\%\\
  \textbf{Baseline+LR} & 79.5\%  & 44.3\% & 63.0\%\\
  \textbf{JOLTW+LR} & \textbf{84.0}\%  & \textbf{51.2}\% & \textbf{66.0}\%\\
  \bottomrule
 \end{tabular}}
\end{center}
 \caption{Part- and bar-wise alignment accuracies for each of our four models over 20 opera versions (12 million frames).}
 \label{tab:experiments}
\end{table}

The Baseline method yields the lowest alignment accuracy. For each alignment, the tracker gets `lost' at the first score jump appearing in the live performance, reflecting the behavior of the algorithm.
The JOLTW model succeeds in tracking accurately the start of each opera version by being able to handle small score jumps when they appear between parts and are included in the 8 part transitions detailed in Section~\ref{sec:HRTracker}. As soon as it faces more severe structural changes, the tracker gets lost and never finds the right score position again. Also, the score position becomes noisy at part transitions. Indeed, the tracker considers the beginning of several close parts as possible candidates before validating one of them.
The Baseline+LR tracker model introduces the LR tracker in the model and shows an important accuracy improvement. In fact, the HR tracker gets lost at each severe jump occurrence but the LR tracker helps retrieve the current score position as soon as it becomes confident in its output position.
The JOLTW+LR tracker, finally, produces the best accuracy in combining the advantages of the two models.

The first column reflects the percentage of performance time that was assigned to the correct part. Each of the two proposed component algorithms -- JOLTW and LR tracker -- improves the global alignment accuracy. The combination of the two gives the best performance, with an accuracy of 84.0\%.
The second column reflects when the tracker is able to highlight the exact bar in the score and the third column indicates the percentage of bars that have an error of up to five bars. Our best model achieves a bar-wise alignment accuracy of 51.2\% and an accuracy of 66.0\% at 5 bars from their ground truth. Even if the results are significantly better than the baseline, the tracker remains not acceptable for a real-life demonstration (e.g., automatic sub-titling), even with our best model. However, we have to remind ourselves that the tests were run on \textit{extremely} mutilated versions of operas (with 33\% to 66\% missing parts and an additional random repetition).

The scores of baseline+LR and JOLTW+LR are close but their alignments reveal different weaknesses. The Baseline+LR model is robust at between-part transitions where jumps do not occur, in following linearly the score, but it has to wait for the correction given by the LR tracker if a jump occurs. JOLTW+LR, on the other hand, has to deal with a set of hypotheses at each part transition, which can tend to add perturbations at part transitions. The accuracy of the JOLTW model reflects how well the adapted JumpDTW algorithm chooses the right hypothesis until the first failure.

A recurrent problem we observed in the experiments is the handling of interludes and jumps appearing before \textit{recitativo} sections. If an interlude appears in this case, the JOLTW algorithm will start the tracking with a musical part. Also, if a jump occurs before restarting at a recitative, the correct recitative selection cannot be reliably selected, due to the very unreliable acoustic similarity value (recitatives generally sound very different between recordings), the LR tracker generally does not retrieve the right position during recitativo sections.

\subsection{Evaluation of the Low-Resolution Tracker}
\label{sec:eval_LR}


This additional experiment is designed to separately test the reliability of the LR tracker by itself. The previous experiment revealed that the LR tracker contributes heavily to improving the results and seems to be a core component of the overall system to assure robustness. In fact, it was designed to consider all score positions during tracking and can retrieve the correct score position when it was not in the scope of the HR tracker or when the latter was lost. The LR dataset involves 400k frames. The results are given in Table~\ref{tab:expe_LR}.

\begin{table} [t]
\small
 \begin{center}
 \scalebox{1}{ 
 \begin{tabular}{@{}lccc@{}}
  \toprule
  \thead{\textbf{{\normalsize Model}}}
    &   \thead{\textbf{{\normalsize Part-wise}}\\ \textbf{{\normalsize align. acc.}}}
    &   \thead{\textbf{{\normalsize Bar-wise}}\\ \textbf{{\normalsize align. acc.}}}
    &   \thead{\textbf{\normalsize {@5 bars}}\\ \textbf{{\normalsize align. acc.}}}\\
  \midrule
  \textbf{LR tracker} & 85.7\%  & 43.5\% & 72.2\%\\
  \textbf{LR track. when rf=1} & 98.9\%  & 50.9\% & 81.7\%\\
  \bottomrule
 \end{tabular}}
\end{center}
 \caption{Part- and bar-wise alignment accuracies for the LR tracker.}
 \label{tab:expe_LR}
\end{table}

The first line reflects the performance of the LR tracker when used as a stand-alone tracker (on the low-resolution dataset). It achieves 85.7\% part-wise and 43.5\% bar-wise accuracy. However, if we limit the evaluation to those passages where the LR reliability factor is set to 1 (which is the situations where the LR tracker can reset the high-resolution tracker), we found that this is the case in 63.5\% of the opera's duration (frames), and there the LR tracker achieves 98.9\% part-wise accuracy and 50.9\% bar-wise alignment accuracy. The near-perfect part-wise accuracy is encouraging, indicating that the LR tracker will retrieve the correct part during the tracking process as soon as its reliability factor is maximal.

\section{Conclusion}
\label{sec:conclusion}

This paper focused on real-time opera tracking, with a focus on the problem of severe structural differences between target and reference. We presented a two-level tracking procedure and demonstrated that it considerably improves tracking robustness, in terms of both  part-wise and bar-wise alignment accuracies. The low-resolution tracker proved to be the core element in this process.
As future work, we are working on extending our opera dataset with new performances (and new operas) where audios and annotations can be freely shared with the research community. Also, a recent machine learning approach that deals with multilingual lyrics-to-audio alignments \cite{vaglio2020multilingual} has shown promising offline results. We will investigate its application for real-time purposes, with a special view on improving the tracking of recitative sections, which are near impossible to precisely align purely at the audio similarity level.

\bibliographystyle{IEEEtran}
\bibliography{IEEEexample}

\end{document}